\documentstyle[12pt,epsf]{article}
 \newcommand{\insertplot}[5]{\begin{figure}
 \hfill\hbox to 0.05in{\vbox to #5in{\vfill
 \inputplot{#1}{#4}{#5}}\hfill}
 \hfill\vspace{-.1in}
 \caption{#2}\label{#3}
 \end{figure}}
 \newcommand{\inputplot}[3]{
 \special{ps: plotfile #1}
}
\newcounter{fixy}
\newenvironment{fixy}[1]{\setcounter{figure}{#1}}
{\addtocounter{fixy}{1}}

\begin{document}

\title{AXIALLY SYMMETRIC MULTISPHALERONS IN YANG-MILLS-DILATON THEORY}
\vspace{1.5truecm}
\author{
{\bf Burkhard Kleihaus, and Jutta Kunz}\\
Fachbereich Physik, Universit\"at Oldenburg, Postfach 2503\\
D-26111 Oldenburg, Germany}

\vspace{1.5truecm}

%\date{September 23, 1996}

\maketitle
\vspace{1.0truecm}

\begin{abstract}
We construct sequences of axially symmetric
multisphaleron solutions in 
SU(2) Yang-Mills-dilaton theory.
The sequences are labelled by a winding number $n>1$.
For $n=1$ the known sequence of spherically symmetric 
sphaleron solutions is obtained.
The solutions within each sequence are labelled
by the number of nodes $k$ of the gauge field functions.
The limiting solutions of the sequences 
correspond to abelian magnetic monopoles 
with $n$ units of charge and energy $E \propto n$.
\end{abstract}
\vfill
\noindent {Preprint hep-th/9609180} \hfill\break
\vfill\eject

\section{Introduction}

Pure SU(2) Yang-Mills theory in $3+1$ dimensions
does not possess static, finite energy solutions.
In contrast, SU(2) Yang-Mills-Higgs (YMH) theory
possesses static, finite energy solutions,
and so do SU(2) Einstein-Yang-Mills (EYM)
and Yang-Mills-dilaton (YMD) theory.

YMH theory with a triplet Higgs field 
contains a stable static, spherically symmetric solution,
the 't Hooft-Polyakov monopole \cite{thooft},
whereas YMH theory with a doublet Higgs field 
contains an unstable static, spherically symmetric solution
\cite{dhn,bog,km}.
This solution represents the electroweak sphaleron
in the limit of vanishing Weinberg angle \cite{km,kkb}.
For large Higgs boson masses the theory contains in addition
a sequence of sphaleron solutions (without parity reflection symmetry)
\cite{kb1,yaffe}.

EYM theory possesses a sequence of unstable static, spherically
symmetric solutions for any finite value of the
coupling constants \cite{bm,strau1,volkov1}.
Within the sequence the solutions are
labelled by the number of nodes $k$ of the
gauge field function.
When the dilaton field is coupled to the system,
the solutions persist in the resulting 
Einstein-Yang-Mills-dilaton (EYMD) theory
\cite{don,lav2,maeda,bizon3,neill,kks3}.
Decoupling gravity leads to YMD theory
with the corresponding sequence of unstable static,
spherically symmetric solutions
\cite{lav1,bizon2}.

In YMH theory, beside the 't Hooft-Polyakov monopole,
there exist multimonopoles with magnetic charge $m=n/g$, 
where $n>1$ is the topological charge or winding number
and $g$ is the gauge coupling constant.
The spherically symmetric 't Hooft-Polyakov monopole
has $n=1$.
Following a pioneering numerical study \cite{rr},
axially symmetric multimonopoles have been
obtained analytically in the Prasad-Sommerfield limit
\cite{forg}. 
In this limit the energy of the multimonopoles satisfies 
the Bogomol'nyi bound
$E=4 \pi n \langle \Phi \rangle /g$.

Similarly, beside the electroweak sphaleron,
which carries Chern-Simons number $N_{CS}=1/2$,
there exist axially symmetric multisphalerons in YMH theory
with Chern-Simons number $N_{CS}=n/2$
\cite{kk}.
So it is natural to ask, whether analogous
axially symmetric solutions exist also
in EYM or YMD theory.

In this letter we construct
multisphaleron solutions and their excitations
in YMD theory.
The appropriate axially symmetric ansatz for the multisphaleron
solutions is analogous to the ansatz in YMH theory
\cite{rr,kk,man,kkb}.
Like the YMH multisphalerons,
the YMD multisphalerons are labelled
by an integer $n$,
which represents a winding number with respect to the azimuthal angle
$\phi$. While $\phi$ covers the full trigonometric circle once,
the fields wind $n$ times around.
For $n=1$, spherical symmetry and the known sequence of
YMD sphaleron solutions are recovered.

For each value of $n$ we find a sequence of axially
symmetric solutions, which can be labelled by the number of 
nodes $k$ of the gauge field functions,
analogous to the spherically symmetric case.
For the limiting solutions,
obtained for $k \rightarrow \infty$,
we give an analytic expression.
The energies of the limiting solutions 
satisfy a Bogomol'nyi type relation $E \propto n$,
which represents an upper bound for the energies
of the solutions of the $n$-th sequence.

In section 2 we briefly review the lagrangian,
discuss the ansatz and present the resulting energy functional.
In section 3 we exhibit the multisphaleron 
solutions with $n \le 4$ and $k \le 4$.
We discuss the limiting solutions in section 4
and present our conclusions in section 5.

\section{\bf Axially symmetric ansatz}

Let us consider the lagrangian of YMD theory
\begin{equation}
{\cal L} =  \frac{1}{2} ( \partial_\mu \Phi \partial^\mu \Phi )
            -\frac{1}{2}e^{2 \kappa \Phi} {\rm Tr}( F_{\mu\nu} F^{\mu\nu})
\   \end{equation}
with dilaton field $\Phi$,
SU(2) gauge field $V_\mu$ and field strength tensor
\begin{equation}
F_{\mu\nu}=\partial_\mu V_\nu-\partial_\nu V_\mu 
            - i g [ V_\mu ,V_\nu ] 
\ , \end{equation}
and the coupling constants $\kappa$ and $g$.
This theory possesses a sequence of static
spherically symmetric sphaleron solutions
labelled by an integer $k$,
which counts the number of nodes of the gauge field function
\cite{lav1,bizon2}.

To obtain static axially symmetric multisphaleron solutions,
we choose the ansatz for the SU(2) gauge fields
analogous to the case of multimonopoles \cite{rr,man} and
electroweak multisphalerons \cite{kk}.
We therefore define a set of orthonormal vectors
\begin{eqnarray}
\vec u_1^{(n)}(\phi) & = & (\cos n \phi, \sin n \phi, 0) \ ,
\nonumber \\
\vec u_2^{(n)}(\phi) & = & (0, 0, 1) \ ,
\nonumber \\
\vec u_3^{(n)}(\phi) & = & (\sin n \phi, - \cos n \phi, 0) 
\ . \end{eqnarray}
and expand the gauge fields 
($V_\mu = V_\mu^a \tau^a /2$) as
\begin{equation}
V_0^a(\vec r) = 0 \ , \ \ \
V_i^a(\vec r) = u_j^{i(1)}(\phi) u_k^{a(n)}(\phi) w_j^k(\rho,z)
\ , \end{equation}
whereas the dilaton field satisfies
\begin{equation}
\Phi(\vec r) = \Phi(\rho,z)
\ . \end{equation}
Invariance under rotations about the $z$-axis
and parity reflections leads to the conditions \cite{rr,kk,kkb}
\begin{equation}
w_1^1(\rho,z)=w_2^1(\rho,z)=w_1^2(\rho,z)=
		   w_2^2(\rho,z)=w_3^3(\rho,z)=0 
\ . \end{equation}

The axially symmetric energy functional
\begin{equation}
E = E_\Phi + E_V =
    \int (\varepsilon_\Phi + e^{2 \kappa \Phi} \varepsilon_V )
	      \, d\phi \, \rho d\rho \, dz
\   \end{equation}
contains the energy densities
\begin{equation}
\varepsilon_\Phi= \frac{1}{2} \left[
 (\partial_\rho \Phi )^2 + (\partial_z \Phi )^2 \right]
\    \end{equation}
and
\begin{eqnarray}
\varepsilon_V & = &\frac{1}{2} \left[
  (\partial_\rho w_3^1 + {1\over{\rho}} ( n w_1^3 + w_3^1 )
	 - g w_1^3 w_3^2 )^2
      +  (\partial_z    w_3^1 + {n\over{\rho}}   w_2^3          
	 - g w_2^3 w_3^2 )^2   
     \right.
\nonumber \\
    & + & \left.
         (\partial_\rho w_3^2 + {1\over{\rho}}   w_3^2          
	 + g w_1^3 w_3^1 )^2
      +  (\partial_z    w_3^2 
	 + g w_2^3 w_3^1 )^2   
      +  (\partial_\rho w_2^3 - \partial_z w_1^3 )^2 
      \right]
\ . \end{eqnarray}
It is still invariant under gauge transformations generated by
\cite{rr,kk}
\begin{equation}
 U= e^{i\Gamma(\rho,z) \tau^i u_3^{i(n)}} 
\ , \end{equation}
where the 2-dimensional scalar doublet
$(w_3^1,w_3^2-n/g\rho)$ transforms with
angle $2 \Gamma(\rho,z)$,
while the 2-dimensional gauge field $(w_1^3,w_2^3)$ transforms
inhomogeneously.
We fix this gauge degree of freedom by choosing the
gauge condition \cite{kk,kkb}
\begin{equation}
\partial_\rho w_1^3 + \partial_z w_2^3 =0 
\ . \end{equation}

Changing to spherical coordinates
and extracting the trivial $\theta$-dependence
(present also for spherically symmetric case $n=1$)
we specify the ansatz further \cite{kk}
\nonumber \\
\begin{eqnarray}
w_1^3(r,\theta) \  & 
= &  \ \ {1 \over{gr}}(1 - F_1(r,\theta)) \cos \theta \ , \ \ \ \ 
w_2^3(r,\theta) \    
= - {1 \over{gr}} (1 - F_2(r,\theta) )\sin \theta    \ ,     
\nonumber \\
w_3^1(r,\theta) \  & 
= & - {{ n}\over{gr}}(1 - F_3(r,\theta) )\cos \theta    \ , \ \ \ \
w_3^2(r,\theta) \    
=  \ \ {{ n}\over{gr}}(1 - F_4(r,\theta) )\sin \theta 
\ . \end{eqnarray}
With $F_1(r,\theta)=F_2(r,\theta)=F_3(r,\theta)=F_4(r,\theta)=w(r)$,
$\Phi(r,\theta)=\varphi(r)$ and $n=1$
the spherically symmetric ansatz of ref.~\cite{bizon2} is recovered.

The above ansatz and gauge choice
yield a set of coupled partial differential equations
for the functions $F_i(r,\theta)$ and $\Phi(r,\theta)$.
To obtain regular solutions with finite energy density
with the imposed symmetries, we take as
boundary conditions for the functions $F_i(r,\theta)$ and $\Phi(r,\theta)$
\begin{eqnarray}
r=0 & :          & \ \ F_i(r,\theta)|_{r=0}=1, 
          \ \ \ \ \ \ \ \ \ i=1,...,4, \ 
              \ \   \partial_r \Phi(r,\theta)|_{r=0}=0,
\nonumber \\
r\rightarrow\infty& :    
	       & \ \ F_i(r,\theta)|_{r=\infty}=F(\infty) , \
               \ \ \ i=1,...,4, 
               \ \ \ \Phi(r,\theta)|_{r=\infty}=\Phi(\infty) ,
\nonumber \\
\theta=0& :      & \ \ \partial_\theta F_i(r,\theta)|_{\theta=0} =0, 
	      \ \ \ \ \ \ i=1,...,4, \ 
	       \ \  \partial_\theta \Phi(r,\theta)|_{\theta=0}=0,
\nonumber \\
\theta=\pi/2& :  & \ \ \partial_\theta F_i(r,\theta)|_{\theta=\pi/2} =0, 
               \ \ \ i=1,...,4, \ 
	       \ \ \  \partial_\theta \Phi(r,\theta)|_{\theta=\pi/2}=0
\ , \label{bc} \end{eqnarray}
(with the exception of $F_2(r,\theta)$ for $n=2$,
which has $\partial_\theta F_2(r,\theta)|_{\theta=0} \ne 0$
\cite{kknew})
where $F(\infty) = \pm 1$ and $\Phi(\infty)=0$.
The boundary conditions for the gauge field functions at infinity
imply, that the solutions are magnetically neutral.
A finite value of the dilaton field at infinity
can always be transformed to zero via
$\Phi \rightarrow \Phi - \Phi(\infty)$, 
$r \rightarrow r e^{-\kappa \Phi(\infty)} $.

The variational solutions
$F_i(\beta r, \theta)$ and $\Phi(\beta r, \theta)$,
lead to $E_\beta = \beta^{-1} E_\Phi + \beta E_V$.
Since the energy functional is minimized for $\beta=1$,
the virial relation \cite{bizon2}
\begin{equation}
E_\Phi = E_V
\   \end{equation}
also holds for general $n$. 

We now remove the dependence on the coupling constants
$\kappa$ and $g$ from the differential equations
by changing to the dimensionless coordinate $x=gr /\kappa$ and 
the dimensionless dilaton function $\varphi = \kappa \Phi$.
The energy then scales with the factor $1/(\kappa g)$.

The dilaton field satisfies asymptotically the relation
\begin{equation}
\lim_{x \rightarrow \infty} x^2 \varphi' = D
\ , \end{equation}
where $D$ is the dilaton charge.
The energy is related to the dilaton charge via
\begin{equation}
E =   \frac{4 \pi}{\kappa g} \lim_{x \rightarrow \infty} x^2 \varphi'
 =   \frac{4 \pi}{\kappa g} D
\ . \label{dil} \end{equation}

\section{\bf Multisphaleron solutions}

We solve the equations numerically,
subject to the boundary conditions eqs.~(\ref{bc}).
To map spatial infinity to the finite value $\bar{x}=1$,
we employ the radial coordinate $\bar{x} = \frac{x}{1+x}$.
The numerical calculations are based on the Newton-Raphson
method. The equations are discretized on a non-equidistant
grid in $\bar{x}$ and an equidistant grid in $\theta$, where
typical grids used have sizes $50 \times 20$ and $100 \times 20$
covering the integration region $0\leq\bar{x}\leq 1$ and
$0\leq\theta\leq\pi/2$.
The numerical error for the functions is estimated to be 
on the order of $10^{-3}$.

The energy density $\varepsilon$, defined by
\begin{equation}
 E= \frac{1}{\kappa g} \int \varepsilon (\vec x)
                       x^2 dx \sin \theta d \theta d\phi
\ , \end{equation}
of the axially symmetric multisphaleron solutions
and their excitations 
has a strong peak on the $\rho$-axis,
while it is rather flat along the $z$-axis.
Keeping $n$ fixed and varying $k$,
we observe, that the ratio of the maximum energy density
$\varepsilon_{\rm max}$
to the central energy density 
$\varepsilon(x=0)$
remains almost constant.
Also, with $n$ fixed increasing $k$, the location of the
peak of the energy density approaches zero exponentially.
On the other hand, with fixed $k$ and increasing $n$
the peak of the energy density moves outward.
The central energy density $\varepsilon(x=0)$
as well as the maximum energy density $\varepsilon_{\rm max}$
are shown in Table~1
for the sequences $n=1-4$ with node numbers $k=1-4$.
Also shown in Table~1 is the energy E.

In the following we exhibit as one example 
the multisphaleron solution for $n=3$ and $k=3$.
In Fig.~1 we show the energy density $\varepsilon$.
In Figs.~2a-d we show the gauge field functions
$F_i$, which go from $F_i(0)=1$ to $F_i(\infty)=-1$,
passing three times zero in any direction.
Finally, in Fig.~3 we show the dilaton function $\varphi$.

\section{Limiting solutions for $k\rightarrow \infty$}

The known sequences of solutions often tend to a simpler
limiting solution. 
For the spherically symmetric YMD sequence
with winding number $n=1$, the limiting solution
for $k \rightarrow \infty$ is given by \cite{bizon2}
\begin{equation}
w_\infty=0 \ , \ \ \ \varphi_\infty=- \ln\left( 1+\frac{1}{x} \right)
\ , \label{lim1} \end{equation}
and describes an abelian magnetic monopole with unit charge,
$m=1/g$.
The gauge field functions $w_k$ approach the limiting
function $w_\infty=0$ nonuniformly, because of the boundary
conditions at the origin and at infinity, where $w_k=\pm 1$.

For the axially symmetric sequences with $n>1$ we observe
a similar convergence with node number $k$.
As for $n=1$, the numerical analysis shows, that
the gauge field functions $(F_i)_k$ tend to the constant value zero
in an exponentially increasing region.
For $(F_i)_\infty=0$,
the set of field equations reduces to a single 
ordinary differential equation for $\varphi_\infty$
\begin{equation}
\left(x^2 \varphi_\infty' \right)' - 
2 e^{2 \varphi_\infty} \frac{n^2}{2 x^2} =0
\ . \label{limn1} \end{equation}
This yields the spherically symmetric limiting solution
\begin{equation}
(F_i)_\infty=0 \ , \ \ \ 
\varphi_\infty= - \ln\left( 1+\frac{n}{x} \right)
\ , \label{limn2} \end{equation}
corresponding to an abelian
magnetic monopole with $n$ units of charge.
Thus the limiting solution of the sequence is charged,
whereas all members of the sequence are magnetically neutral.
This phenomenon
is also observed in EYM and EYMD theory. (A detailed discussion
of the convergence there is given in ref.~\cite{kks3}.)

To demonstrate the convergence of the sequence of
numerical solutions for $\varphi_k$
to the analytic solution $\varphi_\infty$,
we show the functions $\varphi_k$ for $k=1-4$
together with the limiting function $\varphi_\infty$ in Fig.~4 
for $n=3$.
The $k$-th function deviates from the limiting function
only in an inner region, which decreases exponentially with $k$.
The value $\varphi_k(x=0)$ decreases roughly linearly with $k$.

The energy of the limiting solution of the sequence $n$
is given by
\begin{equation}
E = \frac{4 \pi}{\kappa g}
    \int_0^\infty \left[ \frac{1}{2} \varphi_\infty'^2 
                         + e^{2 \varphi_\infty}
\frac{n^2}{2 x^4} \right] x^2 dx = \frac{4 \pi}{\kappa g} n
\ , \label{limn5} \end{equation}
i.~e.~the energy satisfies a Bogomol'nyi type relation,
$E \propto n$.
This relation is in agreement with eq.~(\ref{dil}),
since $D=n$ for the limiting solution.
This limiting value for the energy, eq.~(\ref{limn5}), 
represents an upper bound for each sequence,
as observed from Table~1.
The larger $n$, the slower is the convergence to the limiting solution.

Further details on the solutions and the convergence properties
of the sequences will be given elsewhere \cite{kknew}.

\section{\bf Conclusions}

We have constructed sequences of axially symmetric
multisphaleron solutions in YMD theory.
The sequences are characterized by a winding number $n$,
describing the winding of the fields in the azimuthal angle $\phi$,
while the solutions within each sequences are labelled by
the node number $k$ of the gauge field functions.
For $n=1$ the known spherically symmetric sequence is obtained.

The multisphalerons have a torus-like shape.
The maximum of the energy density occurs on the $\rho$-axis.
With fixed $n$ and increasing $k$ the maximum
moves inward along the $\rho$-axis,
whereas with fixed $k$ and increasing $n$ it
moves outward.

For fixed $n$ and $k \rightarrow \infty$ each sequence approaches
an analytically given limiting solution.
The limiting solution has vanishing gauge field functions
and corresponds to an abelian monopole
with $n$ units of magnetic charge.
Because of the coupling to the dilaton, the energy
of this limiting solution is finite.
It satisfies a Bogomol'nyi type relation
with energy $E \propto n$.

The spherically symmetric YMD sphalerons 
possess fermion zero modes \cite{lav4}.
Since the electroweak sphaleron and
multisphalerons have also 
fermion zero modes \cite{bk,kk,kopen}, 
we expect zero modes to be present also
for the YMD multisphalerons.

The spherically symmetric sequence of $n=1$ YMD sphaleron solutions
can be continued in the presence of gravity,
yielding a corresponding sequence of EYMD solutions,
which still depends on a coupling constant 
\cite{don,lav2,maeda,bizon3,neill,kks3}.
By continuity, we conclude, that the axially symmetric $n>1$ YMD
multisphaleron solutions
also exist in the presence of gravity,
representing axially symmetric EYMD solutions.
In the limit of vanishing dilaton coupling constant,
they should reduce to axially symmetric EYM solutions,
and thus axially symmetric 
generalizations of the Bartnik-McKinnon solutions 
\cite{bm}.
The outstanding question then is, whether there also exist the
corresponding EYM and EYMD black hole solutions.

\vfill\eject

\newpage
\begin{table}
\begin{center}
\begin{tabular}{|cc|crrr|} \hline
 $n$ & $k$      &  $E$       & $\varepsilon(x=0)$& $\varepsilon_{max}$ & $\varphi(x=0)$ \\
 \hline 
 $1$ & $1$      &  $0.804$   & $12.48$            &  $12.48$           &  $-1.711$  \\ 
 $1$ & $2$      &  $0.966$   & $636.2$            &  $636.2$           &  $-3.374$  \\  
 $1$ & $3$      &  $0.994$   & $25352.$           &  $25352.$          &  $-5.154$  \\ 
 $1$ & $4$      &  $0.999$   & $968441.$          &  $968441.$         &  $-6.968$  \\ 
 \hline 
 $2$ & $1$      &  $1.336$   & $0.5457$           &  $1.734$           &  $-1.399$  \\ 
 $2$ & $2$      &  $1.773$   & $7.256$            &  $23.08$           &  $-2.430$  \\  
 $2$ & $3$      &  $1.927$   & $77.25$            &  $251.0$           &  $-3.524$  \\ 
 $2$ & $4$      &  $1.977$   & $840.2$            &  $2674.$           &  $-4.673$  \\ 
 \hline 
 $3$ & $1$      &  $1.800$   & $0.192$            &  $0.92256$         &  $-1.279$  \\ 
 $3$ & $2$      &  $2.482$   & $1.629$            &  $7.552$           &  $-2.110$  \\  
 $3$ & $3$      &  $2.785$   & $10.79$            &  $49.94$           &  $-2.951$  \\ 
 $3$ & $4$      &  $2.913$   & $69.88$            &  $322.4$           &  $-3.837$  \\  
 \hline 
 $4$ & $1$      &  $2.231$   & $0.0941$           &  $0.6541$          &  $-1.209$  \\ 
 $4$ & $2$      &  $3.137$   & $0.6388$           &  $4.132$           &  $-1.943$  \\  
 $4$ & $3$      &  $3.588$   & $3.249$            &  $20.52$           &  $-2.649$  \\ 
 $4$ & $4$      &  $3.808$   & $15.92$            &  $99.91$           &  $-3.387$  \\  
  \hline  
\end{tabular}
\end{center} 
\vspace{1.cm} 

{\bf Table 1}\\
The energy $E$ (in units of $\frac{4 \pi}{\kappa g}$),
the energy density at the origin $\varepsilon(x=0)$,
the maximum of the energy density $\varepsilon_{\rm max}$
and the dilaton field at the origin $\varphi(x=0)$
are shown for the solutions with node numbers $k=1-4$
for the sequences $n=1-4$.
\vspace{1.cm} \\
\end{table}

%XXXXXXXXXXXXXXXXXXXXXX Figure 1 XXXXXXXXXXXXXXXXXXXXXXXXXXXXXXXXX

\newpage
\begin{fixy}{-1}
\begin{figure}
\centering
\epsfysize=11cm
\mbox{\epsffile{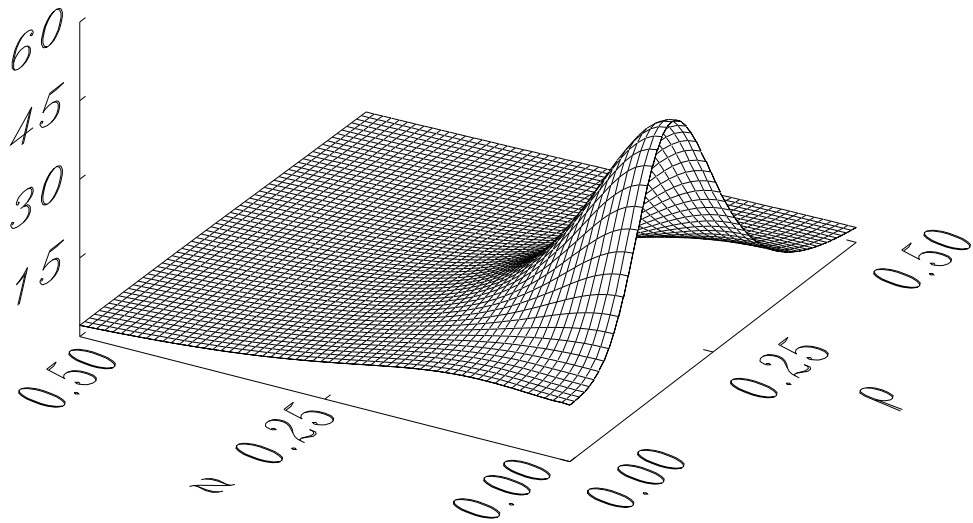}}
\caption{\label{fig1}
The energy density $\varepsilon (\rho,z)$ 
(in units of $\frac{4 \pi}{\kappa g}$) 
is shown as a function of 
the dimensionless coordinates $\rho$ and $z$
for the multisphaleron with  winding number $n = 3$ and 
node number $k = 3$
}
\end{figure}
\end{fixy}
%XXXXXXXXXXXXXXXXXXXXXX Figure 2a XXXXXXXXXXXXXXXXXXXXXXXXXXXXXXX

\newpage
\begin{fixy}{0}
\begin{figure}
\centering
\epsfysize=11cm
\mbox{\epsffile{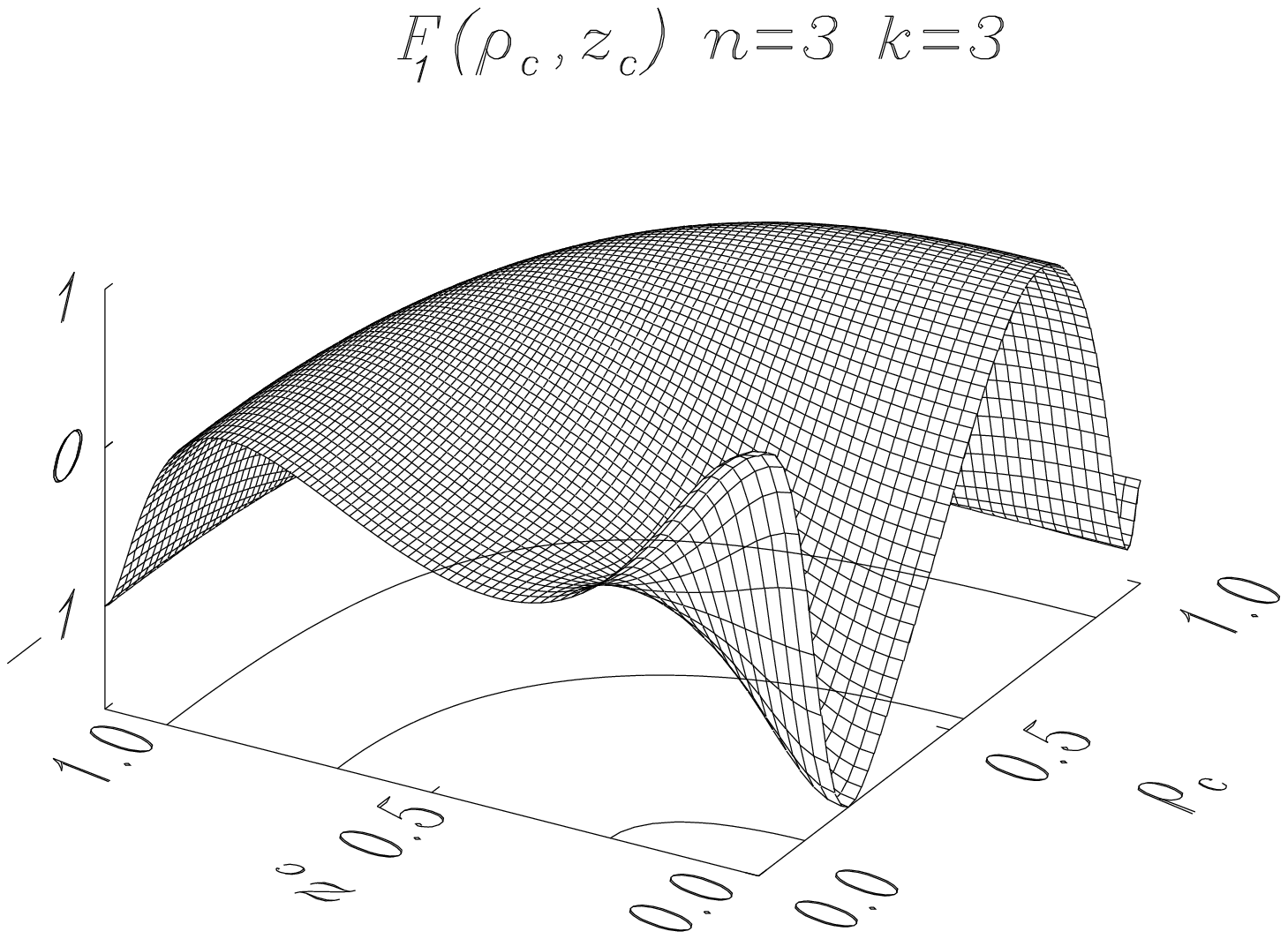}}
\caption{\label{fig2a}
The gauge field function $F_1(\rho_c,z_c)$ 
is shown as a function of 
the dimensionless coordinates $\rho_c$ and $z_c$
for the multisphaleron with  winding number $n = 3$ and 
node number $k = 3$.
The coordinates $\rho_c$ and $z_c$ are defined by 
$\rho_c = \frac{x}{1+x} \sin \theta$ and $z_c = \frac{x}{1+x} \cos \theta$, 
respectively.
The contourlines indicate the nodes of the function.
}
\end{figure}

%XXXXXXXXXXXXXXXXXXXXXX Figure 2b XXXXXXXXXXXXXXXXXXXXXXXXXXXXXXX
\newpage

\begin{figure}
\centering
\epsfysize=11cm
\mbox{\epsffile{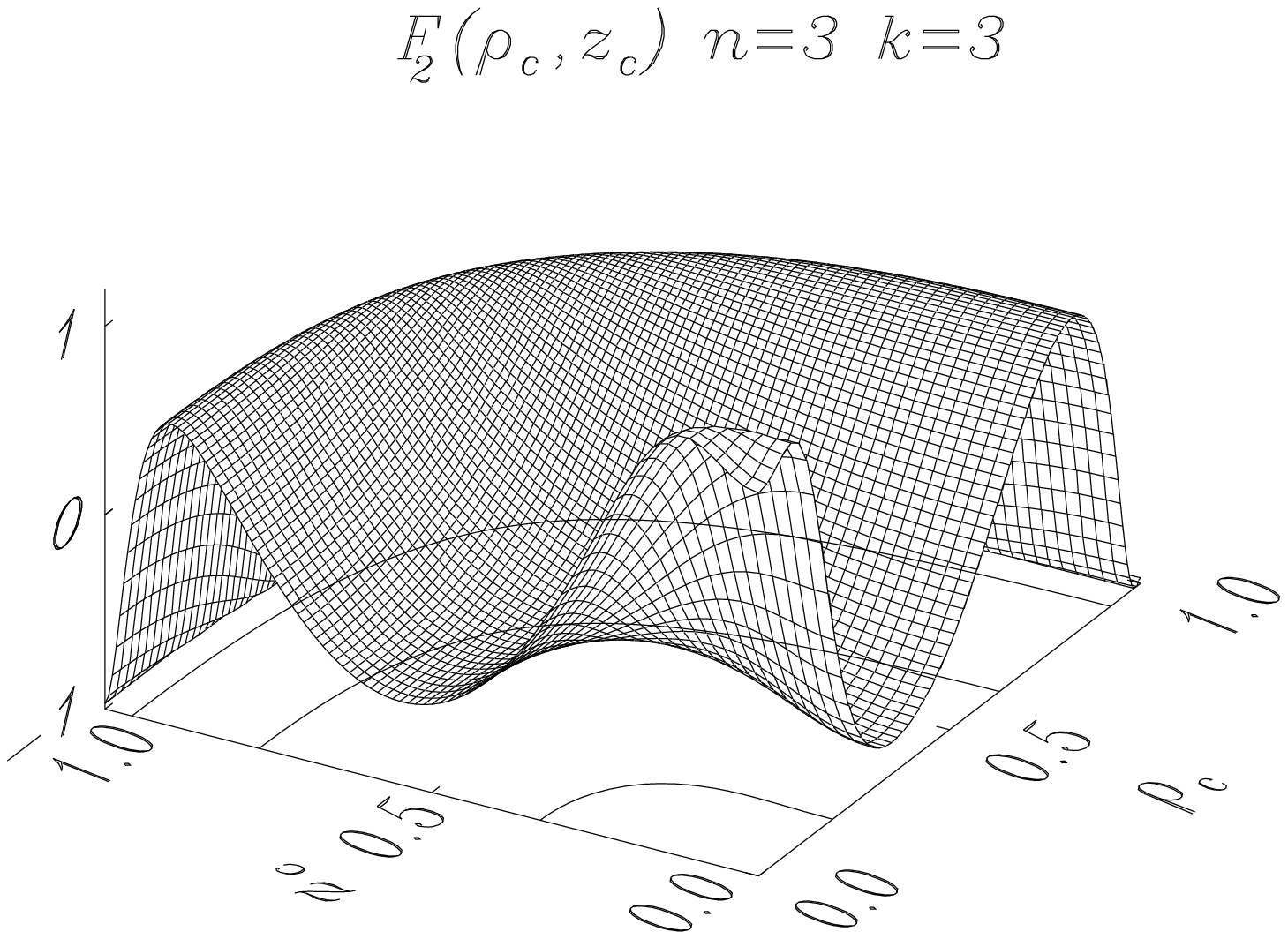}}
\caption{\label{fig2b}
The same as Figure 2a for the gauge field function 
$F_2(\rho_c,z_c)$. 
}
\end{figure}

%XXXXXXXXXXXXXXXXXXXXXX Figure 2c XXXXXXXXXXXXXXXXXXXXXXXXXXXXXXX
\newpage

\begin{figure}
\centering
\epsfysize=11cm
\mbox{\epsffile{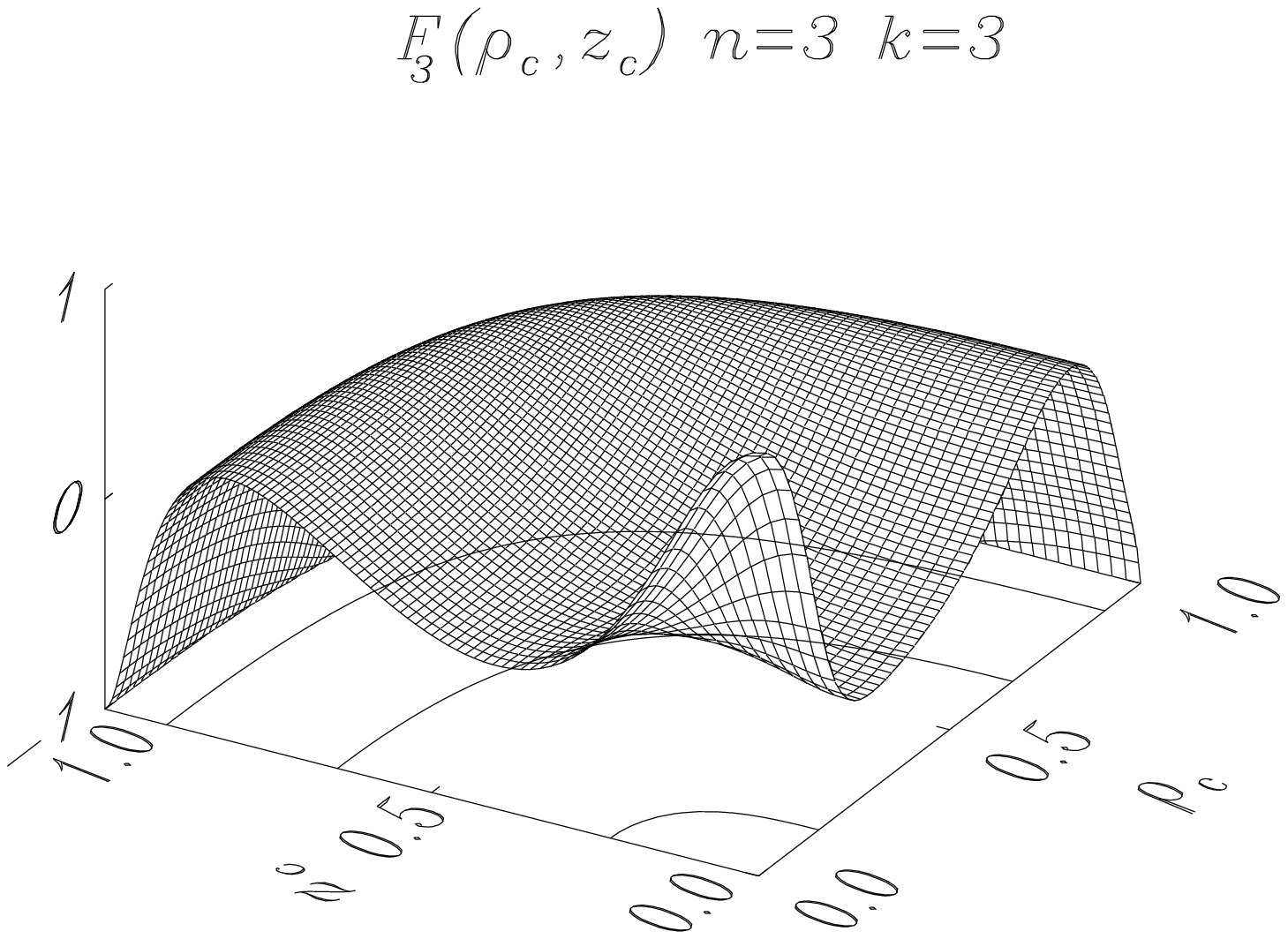}}
\caption{\label{fig2c}
The same as Figure 2a for the gauge field function 
$F_3(\rho_c,z_c)$. 
}
\end{figure}
%XXXXXXXXXXXXXXXXXXXXXX Figure 2d XXXXXXXXXXXXXXXXXXXXXXXXXXXXXXX
\newpage

\begin{figure}
\centering
\epsfysize=11cm
\mbox{\epsffile{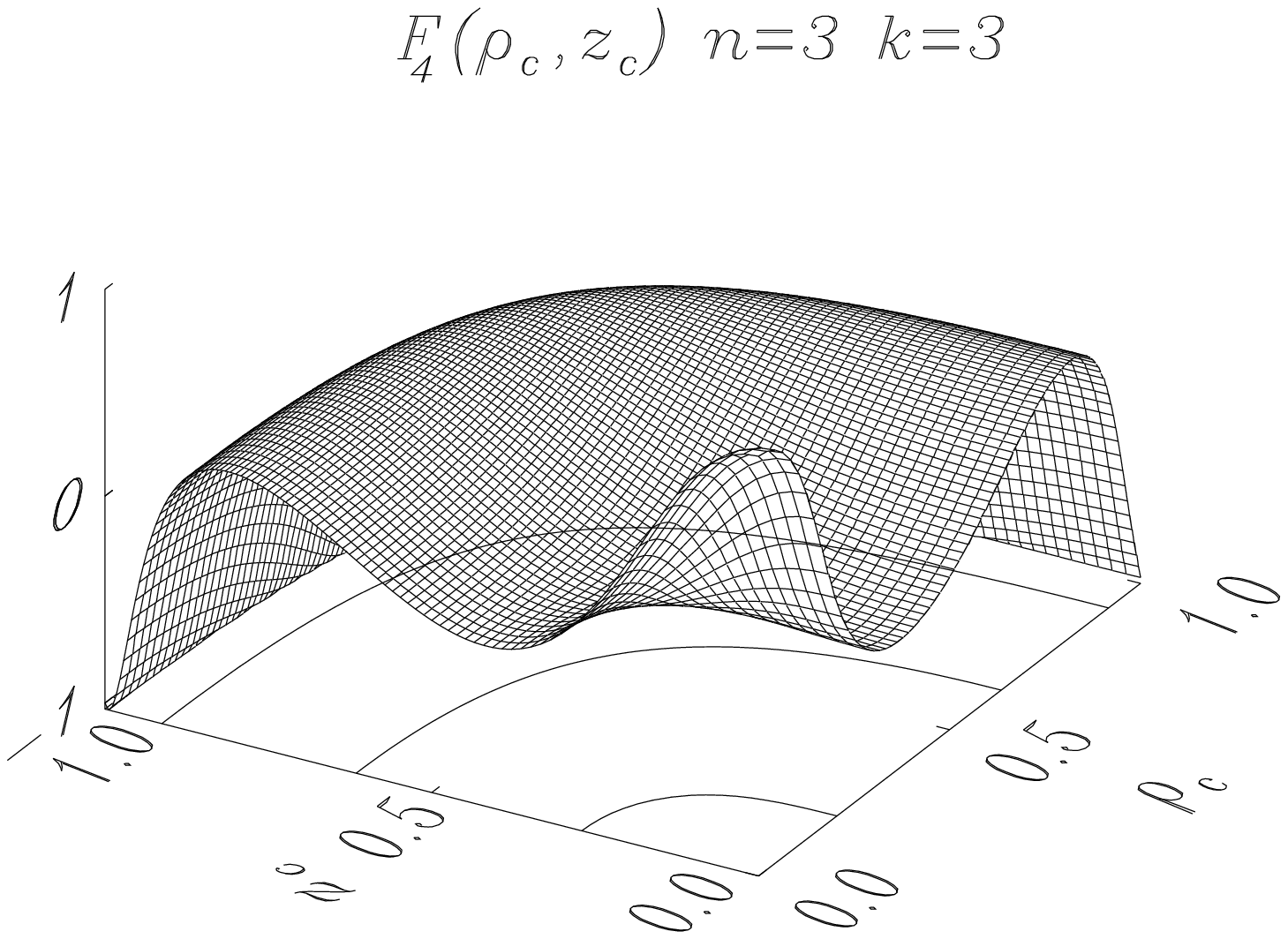}}
\caption{\label{fig2d}
The same as Figure 2a for the gauge field function 
$F_4(\rho_c,z_c)$. 
}
\end{figure}

\end{fixy}

%XXXXXXXXXXXXXXXXXXXXXX Figure 3 XXXXXXXXXXXXXXXXXXXXXXXXXXXXXXXXX

\newpage
\begin{fixy}{-1}
\begin{figure}
\centering
\epsfysize=11cm
\mbox{\epsffile{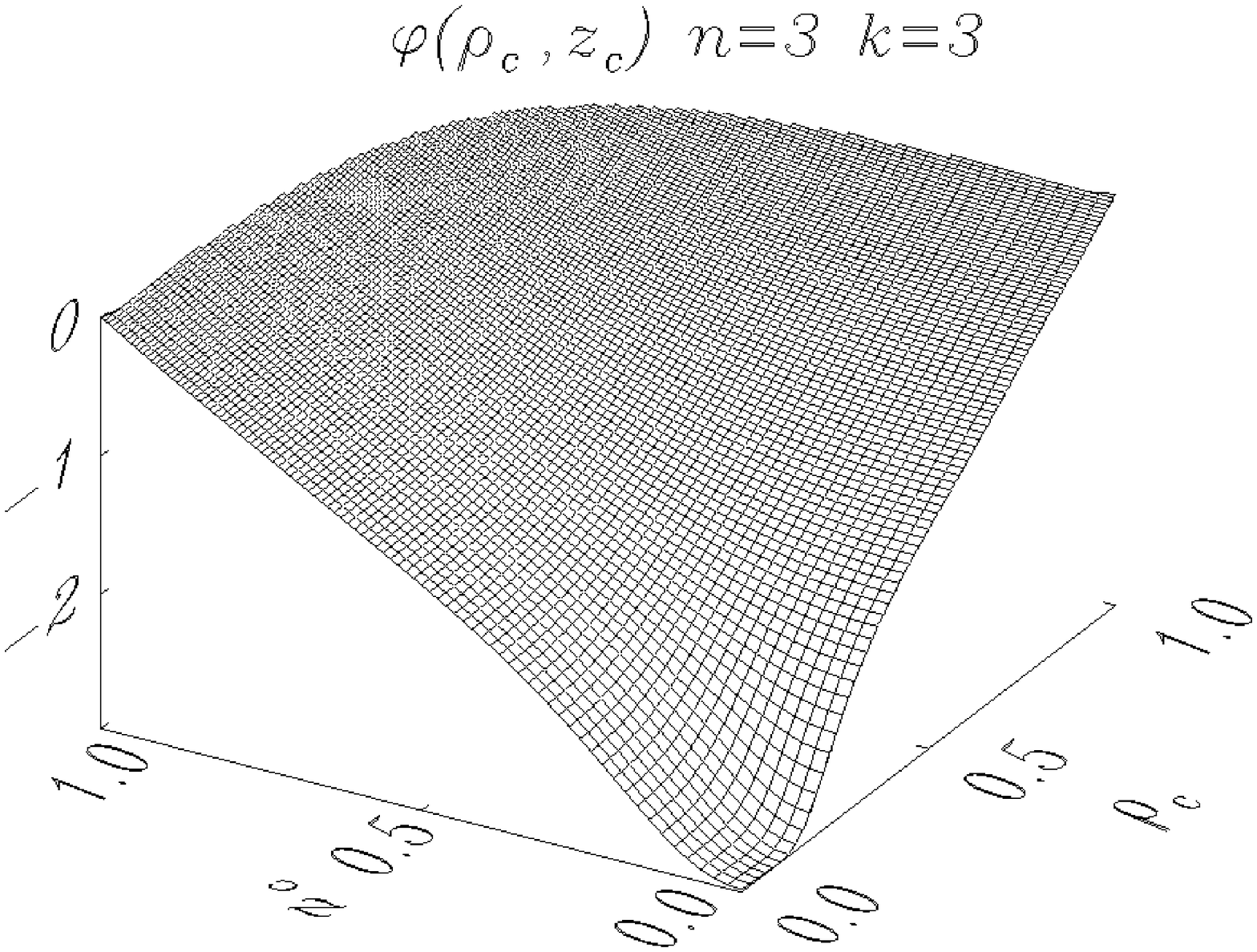}}
\caption{\label{fig3}
The same as Figure 2a for the dilaton function 
$\varphi(\rho_c,z_c)$. 
}
\end{figure}
\end{fixy}

%XXXXXXXXXXXXXXXXXXXXXX Figure 4 XXXXXXXXXXXXXXXXXXXXXXXXXXXXXXXXX

\newpage
\begin{fixy}{-1}
\begin{figure}
\centering
\epsfysize=11cm
\mbox{\epsffile{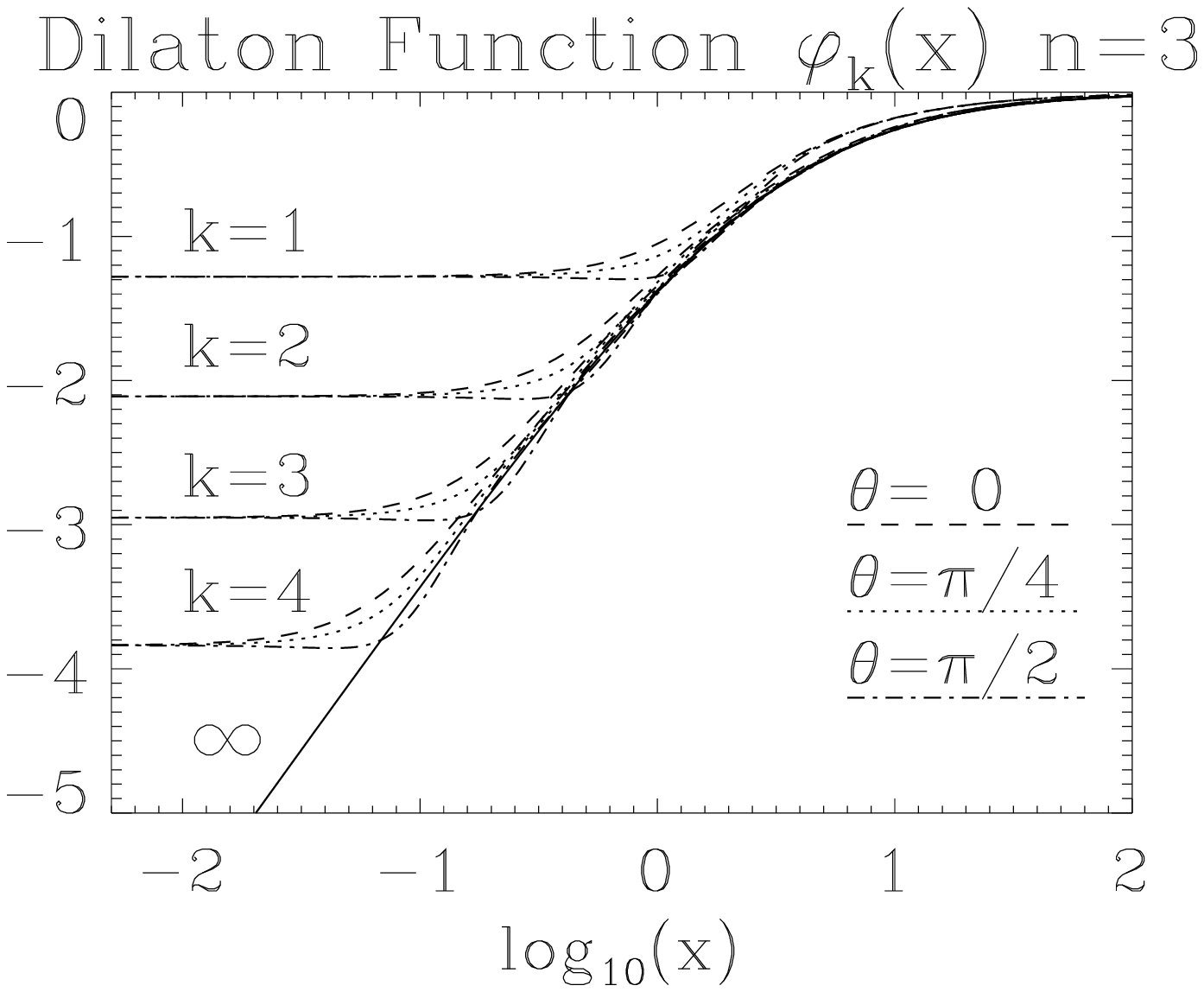}}
\caption{\label{fig4}
The dilaton functions $\varphi_k(x,\theta)$
for the multisphalerons with  winding number $n = 3$ and 
node numbers $k = 1-4$ are shown as a function of the 
dimensionless coordinate $x$. 
The dashed, the dotted and the dash-dotted lines represent the angles 
$\theta = 0$, $\theta = \pi/4$ and $\theta = \pi/2$, respectively.
Also shown is the limiting function $\varphi_\infty (x)$ (solid line). 
}
\end{figure}
\end{fixy}

\end{document}